\documentstyle[twoside,11pt]{article}
\newcommand{\be}{\begin{eqnarray}}
\newcommand{\ee}{\end{eqnarray}}
\newcommand\non {\nonumber}
\newcommand\uh {\frac u 2}
\newcommand\ie {{\it i.e.\ }}

\newcommand\half{\frac 1 2 }
\newcommand\noi{\noindent}
\newcommand\g{\it g}
\newcommand\tr{\mathop{\rm tr}\nolimits}

\begin{document}

\begin{titlepage}
\strut\hfill BONN--TH--95--09, UMTG--185, hep-th/9504085
\vspace{.5in}
\begin{center}
\LARGE Analytical Bethe Ansatz for $A^{(2)}_{2n-1}$, $B^{(1)}_n$,
$C^{(1)}_n$, $D^{(1)}_n$  \\
\LARGE  quantum-algebra-invariant open spin chains\\[.2in]
\large Simone Artz\footnote{Physics Department, Bonn University,
      Nussallee 12, D-53115 Bonn, Germany}\hfill
 Luca Mezincescu${}^{1,}$\footnote{Permanent address: Physics Department,
      University of Miami, Coral Gables, FL 33124 USA}\hfill
 Rafael I. Nepomechie\footnote{Physics Department, University of Miami,
      Coral Gables, FL 33124 USA}
\end{center}
\vspace{.5in}
\begin{abstract}
We determine the eigenvalues of the transfer matrices for integrable open
quantum spin chains which are associated with the affine Lie algebras
$A^{(2)}_{2n-1}$, $B^{(1)}_n$,$C^{(1)}_n$, $D^{(1)}_n$, and which have
the quantum-algebra invariance $U_q(C_n)$, $U_q(B_n)$, $U_q(C_n)$, $U_q(D_n)$,
respectively.

\end{abstract}
\end{titlepage}

\setcounter{footnote}{0}

\section{}

We recently generalized in \cite{our} the analytical Bethe Ansatz for
integrable open spin chains with quantum-algebra invariance which was
developed in \cite{I} to the entire $A^{(2)}_{2n}$ series of
$U_q(B_n)$-invariant spin chains in the fundamental representation.
(The analytical Bethe Ansatz for closed spin chains with periodic boundary
conditions was formulated by Reshetikhin \cite{reshetikhin}.) We focused
in \cite{our} on the $A^{(2)}_{2n}$ series because we sought
to identify the main difficulties in generalizing the analytical Bethe Ansatz
procedure to any affine Lie algebra, and the $A^{(2)}_{2n}$ series was
particularly convenient since the Izergin-Korepin \cite{izergin/korepin}
$A^{(2)}_2$ case was already understood \cite{I}. The main difficulties were
computing the pseudovacuum eigenvalue of the transfer matrix, and formulating
an appropriate Ansatz for general eigenvalues.

In \cite{our} a ``doubling postulate'' (i.e, that the Bethe Ansatz
equations are ``doubled'' with respect to those of the corresponding closed
chain with periodic boundary conditions) was introduced. Using this "doubling
postulate", we were
able to easily formulate an appropriate Ansatz and obtain the transfer matrix
eigenvalues. \footnote{This postulate provides a short-cut for obtaining the
transfer matrix eigenvalues. In principle, this postulate can be avoided, and
the transfer matrix eigenvalues can be obtained by carefully implementing the
constraints of analyticity, crossing, fusion, asymptotic behavior, and
periodicity. In practice, however, this can be quite tedious.} Very recently
this procedure was used for the $G^{(1)}_2$ spin chain by Yung and Batchelor
\cite{yung/batchelor1}. These authors have further generalized this method to
certain open spin chains which are not quantum-algebra-invariant
\cite{yung/batchelor1,yung/batchelor2}.

The eigenvalues of the transfer matrix have been obtained (by means
of the algebraic Bethe Ansatz) also for the $A^{(1)}_n$ open spin chains
\cite{sklyanin,devega/ruiz1}, and the $B^{(1)}_1$ open spin
chain \cite{mnr}.

The success of the analytical Bethe Ansatz procedure
gives us confidence that the same procedure should work for the remaining
series of quantum-algebra-invariant open spin chains, with Hamiltonian
${\cal H} = \sum_{j=1}^{N-1} d \check R_{j,j+1}(u)/du \vert_{u=0} \,.$
Here $R(u)$ is the $R$ matrix associated with an affine Lie algebra
$g^{(k)}$ and with the fundamental representation of $g_0$, where
$g_0 \subset g^{(k)}$ is the maximal finite-dimensional subalgebra of
$g^{(k)}$. Unfortunately, the (diagonal) $K$ matrix which is needed to
construct the corresponding transfer matrix is known \cite{ijmpa}
only for the following additional series of $R$ matrices: $A^{(2)}_{2n-1}$,
$B^{(1)}_n$, $C^{(1)}_n$, $D^{(1)}_n$. We therefore restrict ourselves
to these cases.

Specifically, in this paper we determine the eigenvalues of the transfer
matrices for the following four infinite series of quantum-algebra invariant
open spin chains:

\begin{itemize}

\item $U_q(C_n)$-invariant $A^{(2)}_{2n-1}$ spin chains

\item $U_q(B_n)$-invariant $B^{(1)}_n$  $(n>1)$ spin chains

\item $U_q(C_n)$-invariant $C^{(1)}_n$ spin chains

\item $U_q(D_n)$-invariant $D^{(1)}_n$ spin chains

\end{itemize}

We emphasize that within the class of integrable open spin chains, those
spin chains which are quantum-algebra invariant are in certain respects
the {\it simplest} ones. Indeed, for the quantum-algebra-invariant spin chains,
the Bethe Ansatz states are highest-weight states of the quantum
algebra \cite{pasquier/saleur} - \cite{devega/ruiz2}.
Moreover, for those open spin chains which are not quantum-algebra invariant,
the Bethe Ansatz equations have additional factors depending on some
boundary parameters.

\section{}

We therefore work with solutions $R(u)$ \cite{bazhanov,jimbo1} of the
Yang-Baxter equations which are associated with the affine Lie algebras
$g^{(k)} = (A_{2n-1}^{(2)} \,, B_n^{(1)} \,, C_n^{(1)} \,, D_n^{(1)} )$
and with the fundamental representation of the Lie algebras
$g_0 = (C_n \,, B_n \,, C_n \,, D_n )\,,$ respectively. The corresponding
matrices $\check R(u)$ commute with the generators of the quantum algebras
$U_q(g_0)$.
In the appendix we collect the necessary information about these solutions.
We follow the notations of \cite{our}.

Our goal is to determine the eigenvalues of the transfer
matrix \cite{sklyanin,ijmpa}
\be
   t(u) = \tr_a M_a\ T_a(u)\  \hat T_a(u) \,, \label{transfer}
\ee
where
\be
       T_a(u) &=& R_{aN}(u)\ R_{a N-1}(u)\ \cdots R_{a1}(u)
                      \,,  \non \\
  \hat T_a(u) &=& R_{1a}(u)\ \cdots R_{N-1 a}(u)\ R_{Na}(u)
           \,,  \label{monodromy}
\ee
with the subscript $a$ denoting the auxiliary space while the subscripts
$1, \cdots, N$ refer to quantum spaces. The matrix $M$ is given in
the appendix. It is related to the crossing matrix $V$, $ M = V^t\ V $,
where
\be
    R_{12}(u) = V_1\ R_{12}(-u-\rho)^{t_2}\ V_1
\ee
with
\be
   \rho =\left\{\begin{array}{lll}
                - i \pi -2 \kappa \eta &\mbox{for} & A_{2n-1}^{(2)}\\
                 -2 \kappa \eta &\mbox{for} & B_n^{(1)},\ C_n^{(1)},\ D_n^{(1)}
                 \end{array}\right.
  \label{rho}
\ee
and $\kappa =(2n,\ 2n-1,\ 2n+2,\ 2n-2)$ for
$(A_{2n-1}^{(2)},\ B_n^{(1)},\ C_n^{(1)},\ D_n^{(1)})$, respectively.
The transfer matrix commutes with $U_q(g_0)$ \cite{kulish/sklyanin,mpla}.
We consider simultaneous eigenstates of the
transfer matrix $t(u)$ and the $n$ Cartan generators $\left\{ H_1 \,,
\cdots \,, H_n \right\}$ of $U_q(g_0)$. We call the corresponding eigenvalues
$\Lambda^{(m_1 \,, \cdots \,, m_n)}(u)$ and
$\left\{ \lambda_1  \,, \cdots \,, \lambda_n \right\}$, respectively.
The eigenvalues of the Cartan generators are related to the integers
$m_1 \,, \cdots \,, m_n$ by\footnote{We correct a
typo in \cite{reshetikhin} for the case $C^{(1)}_n$.} \cite{reshetikhin}
\be
 \{\lambda_l\} &=& \{ N - m_1 \,,  m_1 - m_2 \,, \cdots   m_{n-1} - m_n\}
 \hspace{.7cm}\mbox{for}\quad B_n^{(1)} \,, \non
 \\
\{\lambda_l\} &=&\{ N - m_1 \,,   m_1 - m_2 \,, \cdots  \,, m_{n-2} -
m_{n-1} -m_n\,,
  m_{n-1} -m_n\,\} \non \hspace{.7cm}\mbox{for}\quad D_n^{(1)} \,,
 \\
\{\lambda_l\} &=& \{N - m_1 \,,  m_1 - m_2 \,,  \cdots\,,  m_{n-1} -2 m_n \}
\non \hspace{.7cm}\mbox{for}\quad A_{2n-1}^{(2)}\,,C_n^{(1)} \,.
\ee
We choose $\Lambda^{(m_1 \,, \cdots \,, m_n)}(u)$ to correspond to a highest
weight vector for the corresponding algebra $U_q(g_0)$.

To accomplish the analytical Bethe Ansatz program we must specify
the following additional information:
\begin{enumerate}
\item  Crossing relation \cite{I}

\be
    \Lambda^{(m_1 \,, \cdots \,, m_n)}(u) =
    \Lambda^{(m_1 \,, \cdots \,, m_n)}(-u - \rho) \,,
    \label{crossing-lambda}
\ee
with $\rho$ given by Eq. (\ref{rho}).

\item  Fusion formula \cite{I}

\be
    \tilde \Lambda^{(m_1 \,, \cdots \,, m_n)}(u) & = &
    {1\over \alpha(u)^{2N}\ \beta(u)^2}
    \big\{ \zeta( 2u + 2\rho)\ \Lambda^{(m_1 \,, \cdots \,, m_n)}(u)\
    \Lambda^{(m_1 \,, \cdots \,, m_n)}(u+\rho) \non\\
    & & - \zeta( u + \rho)^{2N}\  {\g} (2u + \rho)\ {\g} (-2u - 3\rho) \big\}
    \,.
\label{fusion}
\ee
where
\be
  \zeta(u)=\g(u) \g(-u)
\ee
\be
   \g (u) & = & \left\{\begin{array}{lll}
                2\sinh(\uh+2\eta) \cosh(\uh+\kappa\eta) & \mbox{for}&
                A_{2n-1}^{(2)} \\
                2\sinh(\uh-2\eta) \sinh(\uh+\kappa\eta) &
                \mbox{for}& B_n^{(1)},\ C_n^{(1)},\ D_n^{(1)}
                       \end{array}\right.
  \label{g }
\ee
and
\be
   \alpha(u) &=& \left\{\begin{array}{lll}
                 \cosh(\uh-\kappa\eta) & \mbox{for}&
                  A_{2n-1}^{(2)} \\
                 \sinh(\uh-\kappa\eta) & \mbox{for}&
                  B_n^{(1)},\ C_n^{(1)},\ D_n^{(1)}
                        \end{array}\right.\non\\
    \beta(u) &=&  \sinh(u-2\kappa\eta)
\ee
The fusion formula will be used in order to check the correctness of the
pseudovacuum eigenvalue.

\item  The transfer matrix is periodic $t(u) = t(u+ 2 \pi i)$, and its
       eigenvalues $\Lambda^{(m_1 \,, \cdots \,, m_n)}(u)$ are analytic
       functions of $u$.
\end{enumerate}

To obtain the pseudovacuum eigenvalue of the transfer matrix
we compute its expectation value  in the
pseudovacuum state for $N=2, 3$ and we obtain
\be 
   \Lambda^{(0 \,, \cdots \,, 0)}(u)
&=& c^{2N} \{e^{M_{11} \eta}-\frac{p_0\ {\bar e}^2}{b^2-c^2}
   + e^{M_{mm} \eta}[\frac{a_{m1}^2}{c^2-a_{mm}^2}
   + \frac{{\bar e}^2\ d^2}{(c^2-b^2)(c^2-a_{mm}^2)}]\}\non\\
 &+& b^{2N} \{ p_0 (1 + \frac{{\bar e}^2}{b^2-c^2})
   + e^{M_{mm} \eta}[\frac{ d^2}{b^2-a_{mm}^2}
   + \frac{{\bar e}^2\ d^2}{(b^2-c^2)(b^2-a_{mm}^2)}]\}\non\\
 &+& a_{mm}^{2N} \ e^{M_{mm} \eta}\{ 1+\frac{ d^2}{a_{mm}^2-b^2}
  + \frac{a_{m1}^2}{a_{mm}^2-c^2}
  + \frac{{\bar e}^2\ d^2}{( a_{mm}^2-c^2)( a_{mm}^2-b^2)}\}
\label{pseudo}
\ee
where $p_0 = \sum_{i=2}^{m-1}M_{ii}$, with $M_{ii}$ being the matrix elements
of $M$ given in Eq. (\ref{M}); $d^2 =\sum_{i=2}^{m-1}
a_{mi}^2$; and $a_{\alpha \beta}$, $b$, $c$, and $\bar e$ are given in
Eqs. (\ref{Raa}), (\ref{Ra})  in the appendix. $m= 2n$ for
$A_{2n-1}^{(2)},\ C_n^{(1)},\ D_n^{(1)}$ and $m=2n+1$ for $B_n^{(1)}$.

We postulate that Eq. (\ref{pseudo}) is true for all $N$.
Using Mathematica we find the following expression for the pseudovacuum
eigenvalue

\medskip
\noindent
for $A_{2n-1}^{(2)}$:
\be
 \Lambda^{(0 \,, \cdots \,, 0)}(u) & = & c(u)^{2N}\
                               \frac{\sinh(u-2\kappa\eta)\cosh(u-\omega\eta)}
                                      {\sinh(u-2\eta)\cosh(u-\kappa\eta)}\non\\
                                  &+&b(u)^{2N}\ p_0\
                                  \frac{\sinh(u)\sinh(u-2\kappa\eta)}
{\sinh(u-2\eta)\sinh(u-2(\kappa-1)\eta)}\non\\
                                  &+&a_{mm}(u)^{2N}\
                                   \frac{\sinh(u)\cosh(u-(2\kappa-\omega)\eta)}
                                  {\sinh(u-2(\kappa-1)\eta)\cosh(u-\kappa\eta)}
                                        \label{pseudoA}
\ee

\noindent
for $B^{(1)}_n$, $C^{(1)}_n$, $D^{(1)}_n$:
\be
 \Lambda^{(0 \,, \cdots \,, 0)}(u) & = & c(u)^{2N}\
                                \frac{\sinh(u-2\kappa\eta)\sinh(u-\omega\eta)}
                                    {\sinh(u-2\eta)\sinh(u-\kappa\eta)}\non\\
                                  &+&b(u)^{2N}\ p_0\
                                  \frac{\sinh(u)\sinh(u-2\kappa\eta)}
                                 {\sinh(u-2\eta)\sinh(u-2(\kappa-1)\eta)}\non\\
                                  &+&a_{mm}(u)^{2N}\
                                   \frac{\sinh(u)\sinh(u-(2\kappa-\omega)\eta)}
                                  {\sinh(u-2(\kappa-1)\eta)\sinh(u-\kappa\eta)}
                                        \label{pseudoelse}
\ee
where
\be
p_0=\left\{ \begin{array}{lll}
              \frac{2\sinh(\kappa \eta) \cosh((\kappa - 2)\eta)}{\sinh(2 \eta)}
                     & \mbox{for} & B_n^{(1)} \,, D_n^{(1)}\\ \\
              \frac{2\sinh((2n-2)\eta)\cosh(2n\eta)}{\sinh(2 \eta)}
                      &\mbox{for}&A_{2n-1}^{(2)}, C_n^{(1)}
             \end{array}\right.
\ee
\be
  \omega= \left\{ \begin{array}{lll}
                \kappa -2 & \mbox{for} &  C_n^{(1)}\\
                \kappa +2 &\mbox{for} &A_{2n-1}^{(2)}, B_n^{(1)},
                D_n^{(1)}
                \end{array}\right.
                \label{x}
\ee
and $c(u)$, $b(u)$, $a_{mm}(u)$ are defined in the appendix.
This eigenvalue is consistent with the fusion equation (\ref{fusion}),
\ie the expression for ${\tilde \Lambda}^{(0 \,, \cdots \,, 0)}(u)$
does not have poles for values of $u$ for which $\alpha(u)$ or $\beta(u)$
are zero.

For the dressing of these eigenvalues we make the following Ansatz

\medskip
\noindent
for $A_{2n-1}^{(2)}$:
\be
\Lambda^{(m_1 \,, \cdots \,, m_n)}(u)
         &=& A^{(m_1)}(u)\ c(u)^{2N}
                               \frac{\sinh(u-2\kappa\eta)\cosh(u-\omega\eta)}
                                     {\sinh(u-2\eta)\cosh(u-\kappa\eta)}\non\\
         &+&C^{(m_1)}(u)a_{mm}(u)^{2N}
                                   \frac{\sinh(u)\cosh(u-(2\kappa-\omega)\eta)}
                            {\sinh(u-2(\kappa-1)\eta)\cosh(u-\kappa\eta)}\non\\
         &+& b(u)^{2N}\left\{
         \sum_{l=1}^{n-1} \left[ z_l(u)\ B_l^{(m_l \,, m_{l+1})}(u)
          + \tilde z_l(u)\ \tilde B_l^{(m_l \,, m_{l+1})}(u) \right]
          \right\}\,,
\label{dressedA}
\ee
\noindent
for $B^{(1)}_n$, $C^{(1)}_n$, $D^{(1)}_n$:
\be
   \Lambda^{(m_1 \,, \cdots \,, m_n)}(u)
           & &= A^{(m_1)}(u)\ c(u)^{2N}
                                 \frac{\sinh(u-2\kappa\eta)\sinh(u-\omega\eta)}
                                      {\sinh(u-2\eta)\sinh(u-\kappa\eta)}\non\\
           & &+C^{(m_1)}(u)\ a_{mm}(u)^{2N}
                                   \frac{\sinh(u)\sinh(u-(2\kappa-\omega)\eta)}
                            {\sinh(u-2(\kappa-1)\eta)\sinh(u-\kappa\eta)}\non\\
           & &+ b(u)^{2N}\left \{I w(u)\ B_n^{(m_n)}(u)\   +
              \sum_{l=1}^{n-1} \left[ z_l(u)\ B_l^{(m_l \,, m_{l+1})}(u)
              \right.\right.\non\\
           & & \left.\left.  \qquad\qquad+ \tilde z_l(u)\ \tilde B_l^{(m_l \,,
             m_{l+1})}(u) \right] \right\} \,,
\label{dressedelse}
\ee
where $I=1$ for $B_n^{(1)}$ and $I=0$ in the other cases.
The functions $A, B, C$ are the doubles of the corresponding expressions
given by Reshetikhin \cite{reshetikhin} (apart from slight changes in
notation) and so are also invariant under $u_j^{(l)} \rightarrow -u_j^{(l)}$
\be
  A^{(m_1)}(u)&=&\prod_{j=1}^{m_1}
  \frac{\sinh({1\over 2}({u-u_j^{(1)}})+\eta)\
        \sinh({1\over 2}({u+u_j^{(1)}})+\eta)}
       {\sinh({1\over 2}({u-u_j^{(1)}})-\eta)\
        \sinh({1\over 2}({u+u_j^{(1)}})-\eta)}\,, \\
  \non\\ \non
  C^{(m_1)}(u)&=&A^{(m_1)}(-u-\rho)  \non
\ee
\be
B_l^{(m_l \,, m_{l+1})}(u)&=& \prod_{j=1}^{m_l}
\frac{\sinh({1\over 2}({u-u_j^{(l)}})-(l+2)\eta)\
      \sinh({1\over 2}({u+u_j^{(l)}})-(l+2)\eta)}
     {\sinh({1\over 2}({u-u_j^{(l)}})-l\eta)\
      \sinh({1\over 2}({u+u_j^{(l)}})-l\eta)} \non\\
&\times & \prod_{j=1}^{m_{l+1}}
\frac{\sinh({1\over 2}({u-u_j^{(l+1)}})-(l-1)\eta)\
      \sinh({1\over 2}({u+u_j^{(l+1)}})-(l-1)\eta)}
     {\sinh({1\over 2}({u-u_j^{(l+1)}})-(l+1)\eta)\
      \sinh({1\over 2}({u+u_j^{(l+1)}})-(l+1)\eta)} \,,
      \label{abc} \\
& & \left.\begin{array}{lll}
      l=1\,, \cdots \,, n-2 & \mbox{for}& A_{2n-1}^{(2)}, C_n^{(1)} \\
      l=1\,, \cdots \,, n-1 & \mbox{for}& B_n^{(1)} \\
      l=1\,, \cdots \,, n-3 & \mbox{for}& D_n^{(1)}
      \end{array}\right.  \non
\ee
\be
  \tilde B_l^{(m_l \,, m_{l+1})}(u)&=& B_l^{(m_l \,, m_{l+1})}(-u-\rho) \,,
    \label{Btilde} \non\\
  \tilde z_l(u) &=& z_l(-u-\rho) \,, \qquad\qquad l=1\,, \cdots \,, n-1 \,,
  \label{l}
\ee
\noindent
for $A_{2n-1}^{(2)}$:
\be
B_{n-1}^{(m_{n-1},m_n)}(u)&=&\prod_{j=1}^{m_{n-1}}
\frac{\sinh({1\over 2}({u-u_j^{(n-1)}})-(n+1)\eta)\
      \sinh({1\over 2}({u+u_j^{(n-1)}})-(n+1)\eta)}
     {\sinh({1\over 2}({u-u_j^{(n-1)}})-(n-1)\eta)\
      \sinh({1\over 2}({u+u_j^{(n-1)}})-(n-1)\eta)}\non\\
& & \times \prod_{j=1}^{m_{n}}
\frac{\sinh( u-u_j^{(n)} - 2(n-2)\eta)\
      \sinh( u+u_j^{(n)} - 2(n-2)\eta)}
     {\sinh( u-u_j^{(n)} - 2n\eta)\
      \sinh( u+u_j^{(n)} - 2n\eta)} \,,
\label{ba}
\ee
\noindent
for $B_n^{(1)}$:
\be
B_{n}^{(m_{n})}(u)&=&\prod_{j=1}^{m_{n}}
\frac{\sinh({1\over 2}({u-u_j^{(n)}})-(n-2)\eta)\
      \sinh({1\over 2}({u+u_j^{(n)}})-(n-2)\eta)}
     {\sinh({1\over 2}({u-u_j^{(n)}})-n\eta)\
      \sinh({1\over 2}({u+u_j^{(n)}})-n\eta)}\non\\
& & \times
\frac{\sinh({1\over 2}({u-u_j^{(n)}})-(n+1)\eta)\
      \sinh({1\over 2}({u+u_j^{(n)}})-(n+1)\eta)}
     {\sinh({1\over 2}({u-u_j^{(n)}})-(n-1)\eta)\
      \sinh({1\over 2}({u+u_j^{(n)}})-(n-1)\eta)} \,,
      \label{bb}
\ee
\noindent
for $C_n^{(1)}$:
\be
B_{n-1}^{(m_{n-1},m_n)}(u)&=&\prod_{j=1}^{m_{n-1}}
\frac{\sinh({1\over 2}({u-u_j^{(n-1)}})-(n+1)\eta)\
      \sinh({1\over 2}({u+u_j^{(n-1)}})-(n+1)\eta)}
     {\sinh({1\over 2}({u-u_j^{(n-1)}})-(n-1)\eta)\
      \sinh({1\over 2}({u+u_j^{(n-1)}})-(n-1)\eta)}\non\\
& & \times \prod_{j=1}^{m_{n}}
\frac{\sinh({1\over 2}({u-u_j^{(n)}})-(n-3)\eta)\
      \sinh({1\over 2}({u+u_j^{(n)}})-(n-3)\eta)}
     {\sinh({1\over 2}({u-u_j^{(n)}})-(n+1)\eta)\
      \sinh({1\over 2}({u+u_j^{(n)}})-(n+1)\eta)} \,,
      \label{bc}
\ee
\noindent
for $D_n^{(1)}$:
\be
B_{n-2}^{(m_{n-2},m_{n-1},m_n)}(u)&=&\prod_{j=1}^{m_{n-2}}
\frac{\sinh({1\over 2}({u-u_j^{(n-2)}})-n\eta)\
      \sinh({1\over 2}({u+u_j^{(n-2)}})-n\eta)}
     {\sinh({1\over 2}({u-u_j^{(n-2)}})-(n-2)\eta)\
      \sinh({1\over 2}({u+u_j^{(n-2)}})-(n-2)\eta)}\non\\
     & & \times  \prod_{j=1}^{m_{n-1}}
\frac{\sinh({1\over 2}({u-u_j^{(n-1)}})-(n-3)\eta)\
      \sinh({1\over 2}({u+u_j^{(n-1)}})-(n-3)\eta)}
     {\sinh({1\over 2}({u-u_j^{(n-1)}})-(n-1)\eta)\
      \sinh({1\over 2}({u+u_j^{(n-1)}})-(n-1)\eta)}\non\\
& & \times \prod_{j=1}^{m_{n}}
\frac{\sinh({1\over 2}({u-u_j^{(n)}})-(n-3)\eta)\
      \sinh({1\over 2}({u+u_j^{(n)}})-(n-3)\eta)}
     {\sinh({1\over 2}({u-u_j^{(n)}})-(n-1)\eta)\
      \sinh({1\over 2}({u+u_j^{(n)}})-(n-1)\eta)} \,, \non\\ \non
      \ee
      \be
B_{n-1}^{(m_{n-1},m_n)}(u)&=&\prod_{j=1}^{m_{n-1}}
\frac{\sinh({1\over 2}({u-u_j^{(n-1)}})-(n-3)\eta)\
      \sinh({1\over 2}({u+u_j^{(n-1)}})-(n-3)\eta)}
     {\sinh({1\over 2}({u-u_j^{(n-1)}})-(n-1)\eta)\
      \sinh({1\over 2}({u+u_j^{(n-1)}})-(n-1)\eta)}\non\\
& & \times \prod_{j=1}^{m_{n}}
\frac{\sinh({1\over 2}({u-u_j^{(n)}})-(n+1)\eta)\
      \sinh({1\over 2}({u+u_j^{(n)}})-(n+1)\eta)}
     {\sinh({1\over 2}({u-u_j^{(n)}})-(n-1)\eta)\
      \sinh({1\over 2}({u+u_j^{(n)}})-(n-1)\eta)} \,.
      \label{bd}
\ee

We notice that in contrast to the analytical Bethe Ansatz procedure for closed
spin chains, Eqs. (\ref{dressedA}),(\ref{dressedelse}) contain also the
unknown functions $z_l(u) \,, w(u)$. As mentioned in the introduction, these
functions can be determined by the so called doubling postulate, \ie we
demand that the Bethe Ansatz equations obtained from the cancellation of
poles in $A,B,C$ be ``doubled'' with respect to those in Reshetikhin's
paper \cite{reshetikhin}. The doubled Bethe Ansatz equations are
\be
\left[
\frac{\sinh({u_k^{(1)}\over 2}-\eta)}
     {\sinh({u_k^{(1)}\over 2}+\eta)}\right]^{2N}
&=&\prod_{j \ne k}^{m_1}
\frac{\sinh({1\over 2}({u_k^{(1)}-u_j^{(1)}})-2\eta)\
      \sinh({1\over 2}({u_k^{(1)}+u_j^{(1)}})-2\eta)}
     {\sinh({1\over 2}({u_k^{(1)}-u_j^{(1)}})+2\eta)\
      \sinh({1\over 2}({u_k^{(1)}+u_j^{(1)}})+2\eta)} \non \\
&\times &\prod_{j=1}^{m_2}
\frac{\sinh({1\over 2}({u_k^{(1)}-u_j^{(2)}})+\eta)\
      \sinh({1\over 2}({u_k^{(1)}+u_j^{(2)}})+\eta)}
     {\sinh({1\over 2}({u_k^{(1)}-u_j^{(2)}})-\eta)\
      \sinh({1\over 2}({u_k^{(1)}+u_j^{(2)}})-\eta)} \,, \\
      \non\\ \non\\ \non\\
1&=&  \prod_{j=1}^{m_{l-1}}
\frac{\sinh({1\over 2}({u_k^{(l)}-u_j^{(l-1)}})+\eta)\
      \sinh({1\over 2}({u_k^{(l)}+u_j^{(l-1)}})+\eta)}
     {\sinh({1\over 2}({u_k^{(l)}-u_j^{(l-1)}})-\eta)\
      \sinh({1\over 2}({u_k^{(l)}+u_j^{(l-1)}})-\eta)}\non\\
&\times &  \prod_{j \ne k}^{m_l}
\frac{\sinh({1\over 2}({u_k^{(l)}-u_j^{(l)}})-2\eta)\
      \sinh({1\over 2}({u_k^{(l)}+u_j^{(l)}})-2\eta)}
     {\sinh({1\over 2}({u_k^{(l)}-u_j^{(l)}})+2\eta)\
      \sinh({1\over 2}({u_k^{(l)}+u_j^{(l)}})+2\eta)} \label{BAabcd} \\
& \times & \prod_{j=1}^{m_{l+1}}
\frac{\sinh({1\over 2}({u_k^{(l)}-u_j^{(l+1)}})+\eta)\
      \sinh({1\over 2}({u_k^{(l)}+u_j^{(l+1)}})+\eta)}
     {\sinh({1\over 2}({u_k^{(l)}-u_j^{(l+1)}})-\eta)\
      \sinh({1\over 2}({u_k^{(l)}+u_j^{(l+1)}})-\eta)} \,, \non
\ee
\noindent
where
\be
     \left.\begin{array}{lll}
      l=1\,, \cdots \,, n-1 & \mbox{for}& B_n^{(1)} \\
      l=1\,, \cdots \,, n-2 & \mbox{for}& A_{2n-1}^{(2)}, C_n^{(1)} \\
      l=1\,, \cdots \,, n-3 & \mbox{for}& D_n^{(1)} \,.
      \end{array}\right.
      \label{BAl}
\ee
Moreover, the Bethe Ansatz equations corresponding to values of
$l = 1 \,, 2 \,, \cdots \,, n$ which are not included in Eq. (\ref{BAl})
are as follows:

\medskip
\noindent
for $B_n^{(1)}$:
\be
1&=&   \prod_{j=1}^{m_{n-1}}
\frac{\sinh({1\over 2}({u_k^{(n)}-u_j^{(n-1)}})+\eta)\
      \sinh({1\over 2}({u_k^{(n)}+u_j^{(n-1)}})+\eta)}
     {\sinh({1\over 2}({u_k^{(n)}-u_j^{(n-1)}})-\eta)\
      \sinh({1\over 2}({u_k^{(n)}+u_j^{(n-1)}})-\eta)}\non\\
& \times&  \prod_{j \ne k}^{m_n}
\frac{\sinh({1\over 2}({u_k^{(n)}-u_j^{(n)}})-\eta)\
      \sinh({1\over 2}({u_k^{(n)}+u_j^{(n)}})-\eta)}
     {\sinh({1\over 2}({u_k^{(n)}-u_j^{(n)}})+\eta)\
      \sinh({1\over 2}({u_k^{(n)}+u_j^{(n)}})+\eta)}   \\
      \non
      \label{BAb}
\ee

\noindent
for $A_{2n-1}^{(2)}$:
\be
1&=&   \prod_{j=1}^{m_{n-2}}
\frac{\sinh({1\over 2}({u_k^{(n-1)}-u_j^{(n-2)}})+\eta)\
      \sinh({1\over 2}({u_k^{(n-1)}+u_j^{(n-2)}})+\eta)}
     {\sinh({1\over 2}({u_k^{(n-1)}-u_j^{(n-2}})-\eta)\
      \sinh({1\over 2}({u_k^{(n-1)}+u_j^{(n-2)}})-\eta)}\non\\
& \times&  \prod_{j \ne k}^{m_{n-1}}
\frac{\sinh({1\over 2}({u_k^{(n-1)}-u_j^{(n-1)}})-2\eta)\
      \sinh({1\over 2}({u_k^{(n-1)}+u_j^{(n-1)}})-2\eta)}
     {\sinh({1\over 2}({u_k^{(n-1)}-u_j^{(n-1)}})+2\eta)\
      \sinh({1\over 2}({u_k^{(n-1)}+u_j^{(n-1)}})+2\eta)} \non \\
      & \times&  \prod_{j=1}^{m_n}
\frac{\sinh(u_k^{(n-1)} - u_j^{(n)} + 2\eta)\
      \sinh(u_k^{(n-1)} + u_j^{(n)} + 2\eta)}
     {\sinh(u_k^{(n-1)} - u_j^{(n)} - 2\eta)\
      \sinh(u_k^{(n-1)} + u_j^{(n)} - 2\eta)} \\
            \non\\ \non
      \\
1&=&   \prod_{j=1}^{m_{n-1}}
\frac{\sinh( {u_k^{(n)}-u_j^{(n-1)}}+2\eta)\
      \sinh( {u_k^{(n)}+u_j^{(n-1)}}+2\eta)}
     {\sinh( {u_k^{(n)}-u_j^{(n-1)}}-2\eta)\
      \sinh( {u_k^{(n)}+u_j^{(n-1)}}-2\eta)}\non\\
& \times&  \prod_{j \ne k}^{m_n}
\frac{\sinh( {u_k^{(n)}-u_j^{(n)}}-4\eta)\
      \sinh( {u_k^{(n)}+u_j^{(n)}}-4\eta)}
     {\sinh( {u_k^{(n)}-u_j^{(n)}}+4\eta)\
      \sinh( {u_k^{(n)}+u_j^{(n)}}+4\eta)}  \\
      \non \label{BAa}
\ee
\noindent
for $C_n^{(1)}$:
\be
1&=&   \prod_{j=1}^{m_{n-2}}
\frac{\sinh({1\over 2}({u_k^{(n-1)}-u_j^{(n-2)}})+\eta)\
      \sinh({1\over 2}({u_k^{(n-1)}+u_j^{(n-2)}})+\eta)}
     {\sinh({1\over 2}({u_k^{(n-1)}-u_j^{(n-2}})-\eta)\
      \sinh({1\over 2}({u_k^{(n-1)}+u_j^{(n-2)}})-\eta)}\non\\
& \times&  \prod_{j \ne k}^{m_{n-1}}
\frac{\sinh({1\over 2}({u_k^{(n-1)}-u_j^{(n-1)}})-2\eta)\
      \sinh({1\over 2}({u_k^{(n-1)}+u_j^{(n-1)}})-2\eta)}
     {\sinh({1\over 2}({u_k^{(n-1)}-u_j^{(n-1)}})+2\eta)\
      \sinh({1\over 2}({u_k^{(n-1)}+u_j^{(n-1)}})+2\eta)} \non \\
      & \times&  \prod_{j=1}^{m_n}
\frac{\sinh({1\over 2}({u_k^{(n-1)}-u_j^{(n)}})+2\eta)\
      \sinh({1\over 2}({u_k^{(n-1)}+u_j^{(n)}})+2\eta)}
     {\sinh({1\over 2}({u_k^{(n-1)}-u_j^{(n)}})-2\eta)\
      \sinh({1\over 2}({u_k^{(n-1)}+u_j^{(n)}})-2\eta)} \\
       \non\\ \non
      \\
1&=&   \prod_{j=1}^{m_{n-1}}
\frac{\sinh({1\over 2}({u_k^{(n)}-u_j^{(n-1)}})+2\eta)\
      \sinh({1\over 2}({u_k^{(n)}+u_j^{(n-1)}})+2\eta)}
     {\sinh({1\over 2}({u_k^{(n)}-u_j^{(n-1)}})-2\eta)\
      \sinh({1\over 2}({u_k^{(n)}+u_j^{(n-1)}})-2\eta)}\non\\
& \times&  \prod_{j \ne k}^{m_n}
\frac{\sinh({1\over 2}({u_k^{(n)}-u_j^{(n)}})-4\eta)\
      \sinh({1\over 2}({u_k^{(n)}+u_j^{(n)}})-4\eta)}
     {\sinh({1\over 2}({u_k^{(n)}-u_j^{(n)}})+4\eta)\
      \sinh({1\over 2}({u_k^{(n)}+u_j^{(n)}})+4\eta)}  \\
      \non \label{BAc}
\ee
\noindent
for $D_n^{(1)}$:
\be
1&=&   \prod_{j=1}^{m_{n-3}}
\frac{\sinh({1\over 2}({u_k^{(n-2)}-u_j^{(n-3)}})+\eta)\
      \sinh({1\over 2}({u_k^{(n-2)}+u_j^{(n-3)}})+\eta)}
     {\sinh({1\over 2}({u_k^{(n-2)}-u_j^{(n-3)}})-\eta)\
      \sinh({1\over 2}({u_k^{(n-2)}+u_j^{(n-3)}})-\eta)}\non\\
& \times&  \prod_{j \ne k}^{m_{n-2}}
\frac{\sinh({1\over 2}({u_k^{(n-2)}-u_j^{(n-2)}})-2\eta)\
      \sinh({1\over 2}({u_k^{(n-2)}+u_j^{(n-2)}})-2\eta)}
     {\sinh({1\over 2}({u_k^{(n-2)}-u_j^{(n-2)}})+2\eta)\
      \sinh({1\over 2}({u_k^{(n-2)}+u_j^{(n-2)}})+2\eta)} \non \\
& \times&       \prod_{j=1}^{m_{n-1}}
\frac{\sinh({1\over 2}({u_k^{(n-1)}-u_j^{(n-1)}})+\eta)\
      \sinh({1\over 2}({u_k^{(n-1)}+u_j^{(n-1)}})+\eta)}
     {\sinh({1\over 2}({u_k^{(n-1)}-u_j^{(n-1)}})-\eta)\
      \sinh({1\over 2}({u_k^{(n-1)}+u_j^{(n-1)}})-\eta)}\non\\
& \times&      \prod_{j=1}^{m_{n}}
\frac{\sinh({1\over 2}({u_k^{(n-2)}-u_j^{(n)}})+\eta)\
      \sinh({1\over 2}({u_k^{(n-2)}+u_j^{(n)}})+\eta)}
     {\sinh({1\over 2}({u_k^{(n-2)}-u_j^{(n)}})-\eta)\
      \sinh({1\over 2}({u_k^{(n-2)}+u_j^{(n)}})-\eta)}\\
       \non\\
      \non\\
1&=&   \prod_{j=1}^{m_{n-2}}
\frac{\sinh({1\over 2}({u_k^{(n-1)}-u_j^{(n-2)}})+\eta)\
      \sinh({1\over 2}({u_k^{(n-1)}+u_j^{(n-2)}})+\eta)}
     {\sinh({1\over 2}({u_k^{(n-1)}-u_j^{(n-2)}})-\eta)\
      \sinh({1\over 2}({u_k^{(n-1)}+u_j^{(n-2)}})-\eta)}\non\\
& \times&  \prod_{j \ne k}^{m_{n-1}}
\frac{\sinh({1\over 2}({u_k^{(n-1)}-u_j^{(n-1)}})-2\eta)\
      \sinh({1\over 2}({u_k^{(n-1)}+u_j^{(n-1)}})-2\eta)}
     {\sinh({1\over 2}({u_k^{(n-1)}-u_j^{(n-1)}})+2\eta)\
      \sinh({1\over 2}({u_k^{(n-1)}+u_j^{(n-1)}})+2\eta)} \\
      \non\\ \non
      \\
1&=&   \prod_{j=1}^{m_{n-2}}
\frac{\sinh({1\over 2}({u_k^{(n)}-u_j^{(n-2)}})+\eta)\
      \sinh({1\over 2}({u_k^{(n)}+u_j^{(n-2)}})+\eta)}
     {\sinh({1\over 2}({u_k^{(n)}-u_j^{(n-2)}})-\eta)\
      \sinh({1\over 2}({u_k^{(n)}+u_j^{(n-2)}})-\eta)}\non\\
& \times&  \prod_{j \ne k}^{m_n}
\frac{\sinh({1\over 2}({u_k^{(n)}-u_j^{(n)}})-2\eta)\
      \sinh({1\over 2}({u_k^{(n)}+u_j^{(n)}})-2\eta)}
     {\sinh({1\over 2}({u_k^{(n)}-u_j^{(n)}})+2\eta)\
      \sinh({1\over 2}({u_k^{(n)}+u_j^{(n)}})+2\eta)}  \\
      \non\label{BAd}
\ee

\vspace{0.5cm}

\noi One can therefore determine the unknown functions $z_l(u)$
\be
 z_l(u)& =&\frac{\sinh(u)\sinh(u-2\kappa\eta)\cosh(u-\omega\eta)}
            {\sinh(u-2l\eta)\sinh(u-2(l+1)\eta)\cosh(u-\kappa\eta)}
            \label{za} \\
         & & l=1,\cdots,n-1\qquad\mbox{for}\quad A_{2n-1}^{(2)}\non
\ee
\be
z_l(u)&=&\frac{\sinh(u)\sinh(u-2\kappa\eta)\sinh(u-\omega\eta)}
            {\sinh(u-2l\eta)\sinh(u-2(l+1)\eta)\sinh(u-\kappa\eta)} \\
            \label{zb}
         & & l=1,\cdots,n-1 \qquad\mbox{for}\quad B_n^{(1)},\
            C_n^{(1)},\ D_n^{(1)}\non\\ \non\\
w(u)&=& \frac{\sinh(u)\sinh(u-2\kappa\eta)}
             {\sinh(u- 2n\eta)\sinh(u-2(n-1)\eta)}
             \qquad\mbox{for}\quad B_n^{(1)}
             \label{zc}
\ee

We have obtained expressions for the transfer matrix eigenvalues,
Eqs. (\ref{dressedA}) - (\ref{bd}), (\ref{za}) - (\ref{zc}).
These expressions pass a number of checks which are similar to
those performed in \cite{our} for the $A_{2n}^{(2)}$ case.
We are therefore confident that these are the correct eigenvalues.

\section{}

We conclude by listing some unsolved problems:

The cases which remain to be treated are $D^{(2)}_n$, $D^{(3)}_4$,
and (with the exception of $G^{(1)}_2$) all the exceptional affine
algebras. For these cases, the $R$ and/or $K$ matrices are not yet known.

As noted in the introduction, the analytical Bethe Ansatz method has been
further generalized \cite{yung/batchelor1,yung/batchelor2} to certain open
spin chains which are not quantum-algebra-invariant; namely, spin chains for
which the $K$ matrix is diagonal but is not necessarily equal to the identity
matrix. (For a spin chain with a non-diagonal $K$ matrix, the analytical Bethe
Ansatz method presumably does not work, since an eigenstate (e.g., the
pseudovacuum state) of the transfer matrix is not available.) It would be
interesting to find new diagonal $K$ matrices, and to diagonalize the
corresponding transfer matrices.

Other open problems include formulating the algebraic Bethe Ansatz for
open spin chains (this is known only for the cases $A^{(1)}_n$ and
$B^{(1)}_1$); studying further examples of models with spins in
higher-dimensional representations; and investigating ``graded'' models
associated with solutions of the graded Yang-Baxter equations.
Perhaps the most interesting outstanding problem is to use the Bethe Ansatz
results to investigate boundary phenomena in the thermodynamic
($N \rightarrow \infty$) limit.

\bigskip

We are grateful to  M. Jimbo and A. Kuniba for valuable correspondence.
This work was supported in part by the National Science Foundation
under Grant PHY-92 09978.

\appendix
\section{The $R$ matrix}
The $R$ matrices associated with the fundamental representation
of $ A_{2n-1}^{(2)}$, $B_n^{(1)}$, $ C_n^{(1)}$, $ D_n^{(1)} $
were found by Bazhanov \cite{bazhanov} and
Jimbo \cite{jimbo1}. \footnote{The $A^{(2)}_{2n-1}$ $R$ matrix given in
\cite{bazhanov,jimbo1} is $U_q(D_n)$ invariant.
We consider here a {\it different} $A^{(2)}_{2n-1}$ $R$ matrix, which instead
is $U_q(C_n)$ invariant. We obtain \cite{jimbo2} the latter $R$ matrix from the
$C^{(1)}_n$ $R$ matrix by replacing (in the notation of the first paper
in \cite{jimbo1}) $\xi = k^{2n+2}$ by $\xi = -k^{2n}$; i.e., by changing
$\xi \rightarrow -\xi k^{-2}$. The $R$ matrix obtained in this way
presumably coincides with the $U_q(C_n)$-invariant $A^{(2)}_{2n-1}$ $R$
matrix of Kuniba \cite{kuniba}. We remark that the $A^{(2)}_{2n-1}$ $R$ matrix
given in \cite{jimbo1} can be obtained from the one for $D^{(1)}_n$ by changing
$\xi \rightarrow -\xi k^2$. Similarly, the $A^{(2)}_{2n}$ $R$ matrix,
which is $U_q(B_n)$ invariant, can be obtained from the $B^{(1)}_n$ $R$
matrix by changing $\xi \rightarrow -\xi k^2$.}
We follow the latter reference; however, we use the
variables $u$ and $\eta$ instead of $x$ and $k$, respectively, which
are related as follows:
\be
x = e^u \,, \qquad \qquad k = e^{2 \eta} \,.
\ee
The $R$ matrix is given by \footnote{This expression for the $R$ matrix
differs from the one given in Ref. \cite{jimbo1} by the overall factor
$2 e^{u + (\kappa +2)\eta}$.}
\be
R(u)&=&
c(u)\sum_{\alpha \neq \alpha'}E_{\alpha \alpha} \otimes E_{\alpha \alpha}
+ b(u) \sum_{\alpha \neq\beta,\beta'} E_{\alpha \alpha} \otimes
     E_{\beta \beta} \\
& + & (e(u) \sum_{\alpha < \beta, \alpha \neq \beta'}
+\bar e(u) \sum_{\alpha > \beta,\alpha\neq \beta'})E_{\alpha \beta}
\otimes E_{\beta \alpha}
+\sum_{\alpha \,, \beta} a_{\alpha \beta}(u) E_{\alpha \beta}
\otimes E_{\alpha' \beta'} \non
\ee
with
\be
\left. \begin{array}{lll}
      c(u)&=&2 \sinh(\uh-2\eta)\   \\
      b(u)&=&2 \sinh(\uh)\ \\
      e(u)&=&-2 e^{-\uh} \sinh (2\eta)\  \\
 \bar e(u)&=&e^u e(u) \\
      \end{array}\right \}\ \times
\left \{ \begin{array}{lll}
       \cosh(\uh - \kappa\eta) & \mbox{for} & A_{2n-1}^{(2)} \\
\sinh(\uh -\kappa\eta)  & \mbox{for} & B_{n}^{(1)},\ C_{n}^{(1)},\ D_{n}^{(1)}
       \end{array} \right.
\label{Raa}
\ee
\be
a_{\alpha \beta}(u)=
\left\{ \begin{array}{l}
        2\sinh(\uh)\times\left\{ \begin{array}{l}
     \cosh(\uh -(\kappa-2)\eta)\quad \mbox{for} \quad A_{2n-1}^{(2)} \\
     \sinh(\uh -(\kappa-2)\eta)\quad \mbox{for} \quad B_n^{(1)} \,,
        C_n^{(1)} \,, D_n^{(1)}
                                 \end{array}\right.
              \qquad\qquad \alpha =\beta, \alpha \neq \alpha'\\
              \\
       b(u)-2 \sinh(2\eta)\sinh(2n-1)\eta \quad\mbox{for} \quad B_n^{(1)}
              \qquad\qquad \alpha=\beta, \alpha=\alpha'\\
               \\
       2\sinh 2\eta e^{\mp\uh}\times\left\{ \begin{array}{l}
                      \mp\epsilon_\alpha\epsilon_\beta
                    e^{(\pm \kappa +2(\bar{\alpha}-\bar{\beta}))\eta}\sinh\uh
                        -\delta_{\alpha \beta'} \cosh(\uh-\kappa\eta)
\quad \mbox{for} \quad A_{2n-1}^{(2)} \\
                      \epsilon_\alpha\epsilon_\beta
                    e^{(\pm \kappa +2(\bar{\alpha}-\bar{\beta}))\eta}\sinh\uh
                        -\delta_{\alpha \beta'} \sinh(\uh-\kappa\eta)
\quad \mbox{for} \quad B_n^{(1)} \,, C_n^{(1)} \,, D_n^{(1)}\\
                                  \end{array}\right.
              \quad \alpha _>^< \beta
       \end{array} \right.
\label{Ra}
\ee
where
\be
\kappa=\left\{ \begin{array}{lll}
                2n     & \mbox{for} & A_{2n-1}^{(2)}\\
                2n-1   & \mbox{for} & B_n^{(1)}      \\
                2n+2   & \mbox{for} & C_n^{(1)}      \\
                2n-2   & \mbox{for} & D_n^{(1)}      \\
                \end{array}
                \right.
\ee
and
\be
\bar{\alpha}=\left\{ \begin{array}{ll}
             \alpha-\half & 1\le\alpha \le n\\
             \alpha+\half & n+1\le\alpha\le 2n
             \qquad\mbox{for}\,\,A_{2n-1}^{(2)},\, C_n^{(1)}
                     \end{array} \right.
\ee
\be
\bar{\alpha}=\left\{ \begin{array}{ll}
              \alpha+\half & 1\le\alpha<\frac{{\tilde N}+1}{2}\\
                    \alpha & \alpha=\frac{{\tilde N}+1}{2}\\
              \alpha-\half & \frac{{\tilde N}+1}{2}<\alpha\le {\tilde N}
              \qquad\mbox{for}\,\, B_n^{(1)},\,D_n^{(1)}
                     \end{array} \right.
\ee
\be
   \alpha,\beta &=& 1 \,, \cdots \,, {\tilde N} \non \\
        \alpha' &=& {\tilde N}+1-\alpha
\ee
\be
\epsilon_\alpha & = & \left \{\begin{array}{lll}
                      \ \ 1 & \mbox{for} & 1\le\alpha\le n\\
                       -1 & \mbox{for} & n+1 \le \alpha\le 2n
                        \qquad \mbox{for} \quad A_{2n-1}^{(2)},\,C_n^{(1)}
                               \end{array}\right. \non\\\non\\
\epsilon_\alpha & = &  1 \qquad\qquad \mbox{for}\,\,
B_n^{(1)},\,D_n^{(1)} \\ \non
\ee
where ${\tilde N}=2n\ $ for $ A_{2n-1}^{(2)}$, $ C_n^{(1)}$, $ D_n^{(1)}$ and
${\tilde N}=2n+1$ for $B_n^{(1)}$;
and the $E_{\alpha \beta}$ are elementary matrices.
Evidently, the $R$ matrix acts on the tensor product space
$C^{\tilde N} \otimes C^{\tilde N}$.

In addition to obeying the Yang-Baxter equation, 
this $R$ matrix satisfies the following important properties:

\noi{$PT\ $ {\it symmetry}}
\be
   {\cal P}_{12}\ R_{12}(u)\ {\cal P}_{12} \equiv R_{21}(u)
    = R_{12}(u)^{t_1 t_2} \,;
    \label{pt}
\ee

\noi{\it unitarity }
\be
   R_{12}(u)\ R_{21}(-u)&=&\zeta(u) \,, \label{unitarity}
\ee
where $\zeta (u)$ is given by
\be
 \zeta(u)&=&\left \{ \begin{array}{lll}
             -4\sinh(\uh -2\eta) \cosh(\uh -\kappa\eta)
               \sinh(\uh +2\eta) \cosh(\uh +\kappa\eta) &\mbox{for} &
                 A_{2n-1}^{(2)}\\
             4\sinh(\uh -2\eta) \sinh(\uh -\kappa\eta)
              \sinh(\uh +2\eta) \sinh(\uh +\kappa\eta)  &\mbox{for} &
                 B_n^{(1)},\,C_n^{(1)},\,D_n^{(1)}
               \end{array}\right.
  \label{zeta}
\ee

\noi{\it crossing symmetry}
\be
   R_{12}(u)=V_1\ R_{12}(-u-\rho)^{t_2}\ V_1
   = V_2^{t_2}\ R_{12}(-u-\rho)^{t_1}\ V_2^{t_2} \,,
\label{cross}
\ee
where $\rho=-i \pi -2 \kappa\eta$  for $ A_{2n-1}^{(2)}$ and $\rho=-2
\kappa\eta$ for $B_n^{(1)},\,C_n^{(1)},\,D_n^{(1)}$; and $V^2 = 1$;

\noi{\it regularity}
\be
   R(0)= -\zeta(0)^\half {\cal P}  \,,
\ee
where ${\cal P}$ is the permutation operator
\be
  {\cal P}=\sum_{\alpha, \beta} E_{\alpha \beta}\otimes E_{\beta \alpha}
   \,;
\ee

\noi{\it commutativity}
\be
   \left[ \check{R}(u) \,, \check{R}(v) \right] = 0,
   \hspace{2cm} \check{R} = \cal{P} R \,; \label{Rcheck}
\ee

\noi{\it periodicity}
\be
   R(u + 2\pi i) = R(u)  \,.
\ee

The crossing matrix is given by
\be
   V &=&  E_{\alpha \alpha} \delta_{\alpha \alpha'}
         -\sum_{\alpha\ne \alpha' } e^{({\bar \alpha}-{\bar {\alpha'}})\eta}
         E_{\alpha
         \alpha'}\,.
\label{V}
\ee
Correspondingly, $M=V^t\ V$ is given by the ${\tilde N}\times {\tilde N}$
diagonal matrix
\be
  M & =& diag \left( e^{2(2n)\eta} \,, e^{2(2n-2)\eta}\,, \cdots \,,
e^{4\eta} \,, e^{-4\eta} \,, \cdots \,, e^{-2(2n-2)\eta} \,,
e^{-2(2n)\eta} \right)
\quad \mbox{for} \quad A_{2n-1}^{(2)}, C_n^{(1)} \,,\non\\
  M & =& diag \left( e^{2(2n-1)\eta} \,, e^{2(2n-3)\eta} \,, \cdots \,,
e^{2\eta} \,,1\,, e^{-2\eta} \,, \cdots \,, e^{-2(2n-3)\eta} \,,
e^{-2(2n-1)\eta} \right) \quad \mbox{for} \quad B_n^{(1)} \,,\\
  M & =& diag \left( e^{2(2n-2)\eta} \,, e^{2(2n-4)\eta} \,, \cdots \,,
1 \,, 1 \,, \cdots \,, e^{-2(2n-4)\eta} \,,
e^{-2(2n-2)\eta} \right) \quad \mbox{for} \quad  D_n^{(1)}   \,.\non
\label{M}
\ee

\vfill\eject
\end{document}